\documentclass[12pt]{iopart}
\usepackage{hyperref}
\usepackage{amssymb}
\usepackage[normalem]{ulem}
\usepackage{bm,url}
\usepackage[export]{adjustbox}
\usepackage{color}
\usepackage[usenames,dvipsnames]{xcolor}
\usepackage{graphicx}
\usepackage{lscape}
\newcommand{\psr}{PSR~J0337+1715}
\newcommand{\lpi}{local position invariance}

\begin{document}

\title[Local Position Invariance with \psr{}]{An independent test on the \lpi{}
of gravity with the triple pulsar \psr{}}
\author{Lijing Shao}
\address{Max Planck Institute for Gravitational Physics (Albert Einstein
Institute), \\ Am M\"uhlenberg 1, D-14476 Potsdam-Golm, Germany}
\ead{lijing.shao@aei.mpg.de}

\vspace{10pt}
\begin{indented}
\item[]\today
\end{indented}

\begin{abstract}
We design a direct test of the \lpi{} (LPI) in the post-Newtonian gravity, using
the timing observation of the triple pulsar, \psr{}. The test takes advantage of
the large gravitational acceleration exerted by the outer white dwarf to the
inner neutron star -- white dwarf binary. Using machine-precision three-body
simulations and dedicated Markov-chain Monte Carlo (MCMC) techniques with
various sampling strategies and noise realizations, we estimate that the
Whitehead's parameter could have already been limited to $|\xi| \lesssim 0.4$
(95\% CL), with the published timing data spanning from January 2012 to May
2013. The constraint is still orders of magnitude looser than the best limit,
yet it is able to independently falsify Whitehead's gravity theory where
$\xi=1$.  In addition, the new test is immune to extra assumptions and involves
full dynamics of a three-body system with a strongly self-gravitating neutron
star.
\end{abstract}

\vspace{2pc}
\lpi{}

\submitto{Classical and Quantum Gravity}
\maketitle

%-------------------------------------------------------------------------------
\section{Introduction}
\label{sec:intro}
%-------------------------------------------------------------------------------

\begin{figure}
  \includegraphics[width=13cm,right]{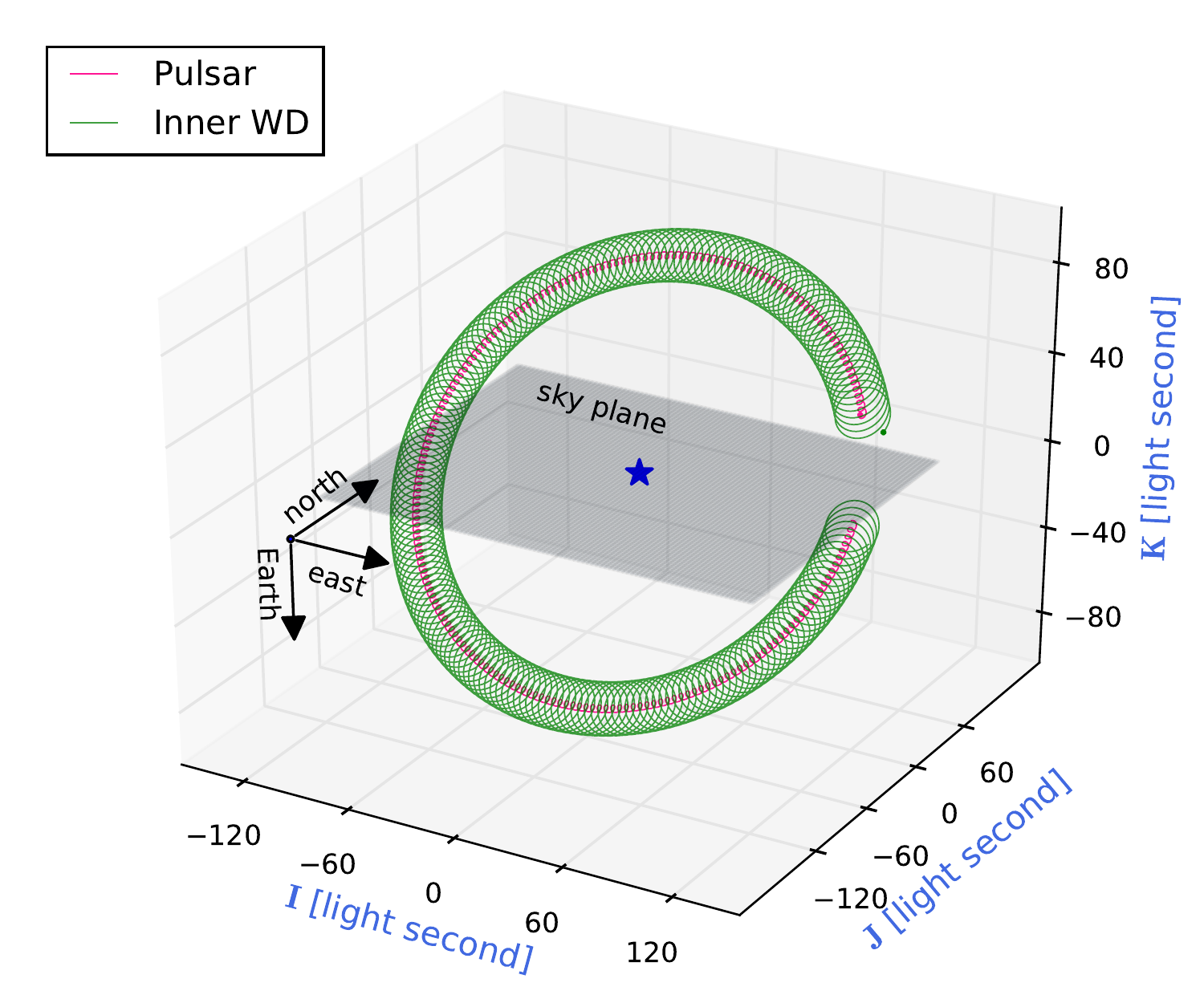}
  \caption{Illustration of the triple pulsar system in the $(\hat{\bf I},
  \hat{\bf J}, \hat{\bf K})$ coordinate system where, the origin (marked as a
  blue star) is chosen to be the center of inertial masses of the triple system,
  $\hat{\bf K}$ points from the Earth to \psr{}, $\hat{\bf I}$ and $\hat{\bf J}$
  respectively point to east and north in the sky plane. The longitude of
  ascending node for the outer orbit, which is generally not an observable in
  pulsar timing, is assumed to be $\Omega_{\rm O} = 0^\circ$; for other nonzero
  values, a rotation around $\hat{\bf K}$ for an angle $\Omega_{\rm O}$ is
  needed.  The trajectories of the pulsar (in pink) and the inner WD (in green)
  start on MJD~55920.0 (December 25, 2011), and end on MJD 56233.9 (November 2,
  2012) when the pulsar was ascending. The starting locations are indicated by
  small dots. The orbit of the outer WD is not shown here.  These trajectories
  are integrated with the {\sc rebound} package (see section~\ref{sec:pe}).
  \label{fig:orbit3d}}
\end{figure}

\psr{}, a millisecond pulsar (MSP) with a spin period $P\simeq 2.73$\,ms, was
discovered in a large-scale pulsar survey conducted with the Robert C. Byrd
Green Bank Telescope (GBT) \cite{rsa+14}. It is in a hierarchical triple
system consisting of a neutron star (NS) with a gravitational mass $m_{\rm NS}
\simeq 1.44\,M_\odot$, and two white dwarfs (WDs) with masses $m_{\rm WD,I}
\simeq 0.20\,M_\odot$ and $m_{\rm WD,O} \simeq 0.41\,M_\odot$.\footnote{In
  this paper, ``\psr{}'' , ``the triple pulsar'', and ``the neutron star'' are
used to refer to the pulsar, while ``the triple system'' and ``the pulsar
system'' are used to refer to the three-body system.} The NS and the lighter
WD are gravitationally bound as an {\it inner binary} with $P_{\rm b,I} \simeq
1.63$\,d that are, as a whole hierarchically bound to the outer WD with $P_{\rm
b,O} \simeq 327$\,d. Two orbits are very circular with $e_{\rm I} \simeq 6.9
\times 10^{-4}$ for the inner binary, and $e_{\rm O} \simeq 3.5 \times 10^{-2}$
for the outer orbit. Two orbital planes are remarkably coplanar with an
inclination $\lesssim 0.01^\circ$ due to the three-body dynamics in the
formation of the system \cite{tv14}. The apsides of two orbits are aligned with
a difference $\lesssim2^\circ$ due to the secular effects of the three-body
interaction \cite{rsa+14}.  The 3-dimensional spatial trajectory of the inner
binary for a time span slightly shorter than $P_{\rm b,O}$  is illustrated in
Figure~\ref{fig:orbit3d}.

%-------------------------------------------------------------------------------
%-------------------------------------------------------------------------------

The triple pulsar is identified immediately as a superb celestial laboratory to
test the strong equivalence principle (SEP) by investigating the difference in
the inertia and gravitational masses of the pulsar \cite{rsa+14}. Later on such
a test is extended to probe the equivalence between the passive and active
gravitational masses as well, that will represent the first test of Newton's
third law with compact objects \cite{sha16,bon57}. Indeed, SEP is the founding
principle of general relativity (GR) that deserves the strictest examination
from every angle, to establish its precision as well as to look for new physics
beyond GR.  Will~\cite{wil93,wil14a} summarized the equivalence principles in
gravity theories in a hierarchical way, from the weak equivalence principle
(WEP) to the Einstein equivalence principle (EEP), and then to SEP, where the
last one can be decomposed into three parts,
\begin{itemize}
  \item the universality of free fall (UFF),
  \item the local Lorentz invariance (LLI),
  \item the local position invariance (LPI),
\end{itemize}
for non-self-gravitating bodies as well as for self-gravitating bodies.  SEP
describes the general rules for the outcome of gravitational experiments. It is
indeed lying to the heart of GR, and actually, there are arguments that
among the viable gravity theories, GR is the only one that respects SEP in its
entirety \cite{wil14a}. Therefore, probing the building blocks of SEP probes
the deepest foundational principle of GR \cite{sw16}.

The tests in the equivalence of different masses pertains to tests of UFF
\cite{rsa+14,sha16}.  In this paper we will study the possibility of using
\psr{} to test the LPI in gravity, which is another important ingredient of
SEP.  The motivation of the study is similar to the tests of UFF. The outer WD
provides a substantial gravitational environment for the inner binary that is
valuable to study UFF and LPI in gravity theories
\cite{fkw12,wex14,bbc+15a,sw16}. In post-Newtonian gravity, the tests of UFF
and LPI with self-gravitating bodies were done with the Sun-Earth-Moon system
\cite{wtb12}, and with the Milky Way -- binary-pulsar systems
\cite{sfl+05,zsd+15}.  These studies would benefit greatly if the third body
were exerting a larger gravitational effect on the other two bodies.
The triple system precisely provides an ideal realization
of this requirement in Nature. In addition, there is a strongly
self-gravitating body involved (namely, the NS), some strong-field aspects
could be studied, which is nearly impossible with weakly self-gravitating
bodies alone (e.g. in the lunar laser ranging experiments~\cite{wtb12}).
Moreover, compared with tests using the Milky Way -- binary-pulsar systems,
the triple system probes a {\it dynamical} regime where
the third body (namely the outer WD) does {\it react} to the gravitational
dynamics. Although the limit on LPI violation that could have been provided
  by the triple pulsar is found to be orders of magnitude looser than the best
  limit from solitary pulsars~\cite{sw13,wil14a}, we still feel it worthy as an
  {\it independent} limit and it has various extra merits compared with
  previous studies (see discussions in section~\ref{sec:dis}).

The paper is organized as follows. In the next section, we present some
theoretical details for the orbital dynamics of \psr{} in presence of LPI
violation, in the parameterized post-Newtonian (PPN) gravity. Then in
section~\ref{sec:pe} we simulate various mock time-of-arrival (TOA) data
closely following the real observation in Ref.~\cite{rsa+14} (see
  Table~\ref{tab:psr}) with a machine-precision $N$-body integrator. Dedicated
  parameter estimation using Markov-chain Monte Carlo (MCMC) techniques that
  take account of full correlations (of 17 parameters) is performed on these
  mock TOAs, and from MCMC chains we conservatively estimate the precision of
  the test.  Section~\ref{sec:dis} discusses the relevance of the results, and
  makes some comparisons with limits obtained elsewhere.

%-------------------------------------------------------------------------------
\section{Triple pulsar system with LPI violation}
\label{sec:lpiv}
%-------------------------------------------------------------------------------

PPN formalism is the most popular framework in experimental gravity for testing
alternative gravity theories. In PPN gravity theories are parametrized with ten
generic PPN parameters that represent different physical properties of
gravitation and take different values in different theories
\cite{wn72,nw72,wil93}. There also exists a generic framework called the
  standard-model extension (SME)~\cite{kos04,bk06} which, using an effective
  field theory, casts the possible deviations from GR into new operators that
  can host anisotropic terms as well. Due to the large number of coefficients
  in SME, here we will focus on the LPI-violating PPN parameter,
  $\xi$. It is also called the Whitehead's parameter
  because of its first appearance in Whitehead's parameter-free gravity theory
  \cite{whi22,wil73,gw08}.  In GR, $\xi=0$, and in Whitehead's gravity theory
  $\xi=1$. The original Whitehead's gravity theory~\cite{whi22} was
  disproved by many experiments by now~\cite{wil73, sw13, wil14a} (see
  Ref.~\cite{gw08} for an excellent review), but due to the importance of the
  SEP it does not hurt to have another {\it independent} test with distinct
merits.

In the PPN formalism, one has an extra LPI-violating term in the $N$-body
Lagrangian \cite{wil93,sw13},
\begin{equation}\label{eq:xi:lagrangian}
    L_\xi = - \frac{\xi}{2} \frac{G^2}{c^2} \sum_{i,j}
    \frac{m_i m_j}{r_{ij}^3} \, \bm{r}_{ij} \cdot
    \left[
    \sum_k m_k \left(
    \frac{\bm{r}_{jk}}{r_{ik}} - \frac{\bm{r}_{ik}}{r_{jk}}
    \right)
    \right] \,,
\end{equation}
where $\bm{r}_{ij} \equiv \bm{r}_i - \bm{r}_j$, $r_{ij} \equiv |\bm{r}_{ij}|$,
and the summation excludes terms that make any denominator vanish.  In the
Hamiltonian formalism, for a triple system  like \psr{} where finite-size
effects are subdominant, $L_\xi$ corresponds to an extra term in the interaction
potential,
\begin{equation}
    \hspace{-2.5cm}
    V_\xi = \xi \frac{G^2}{c^2} m_0 m_1 m_2
    \left(
    \frac{\bm{r}_{01} \cdot \bm{r}_{12}}{r_{01}^3 r_{20}} +
    \frac{\bm{r}_{01} \cdot \bm{r}_{12}}{r_{12}^3 r_{20}} +
    \frac{\bm{r}_{01} \cdot \bm{r}_{20}}{r_{01}^3 r_{12}} +
    \frac{\bm{r}_{01} \cdot \bm{r}_{20}}{r_{20}^3 r_{12}} +
    \frac{\bm{r}_{12} \cdot \bm{r}_{20}}{r_{12}^3 r_{01}} +
    \frac{\bm{r}_{12} \cdot \bm{r}_{20}}{r_{20}^3 r_{01}}
    \right) \,,
    \label{eq:xi:potential}
\end{equation}
where in our case we use subscripts, 0, 1, 2, refer to the NS, the inner WD,
and the outer WD, respectively. Due to the hierarchical structure for the
  triple system that we are considering, we have $ r_{01}  \ll r_{12}  \sim
  r_{20}$. Therefore, the first and the third terms in the parentheses
  contribute prominently to $V_\xi$. Nevertheless, we include all contributions
  in Eq.~(\ref{eq:xi:potential}) in our numerical study below.

%-------------------------------------------------------------------------------
\begin{figure}
  \includegraphics[width=13cm,right]{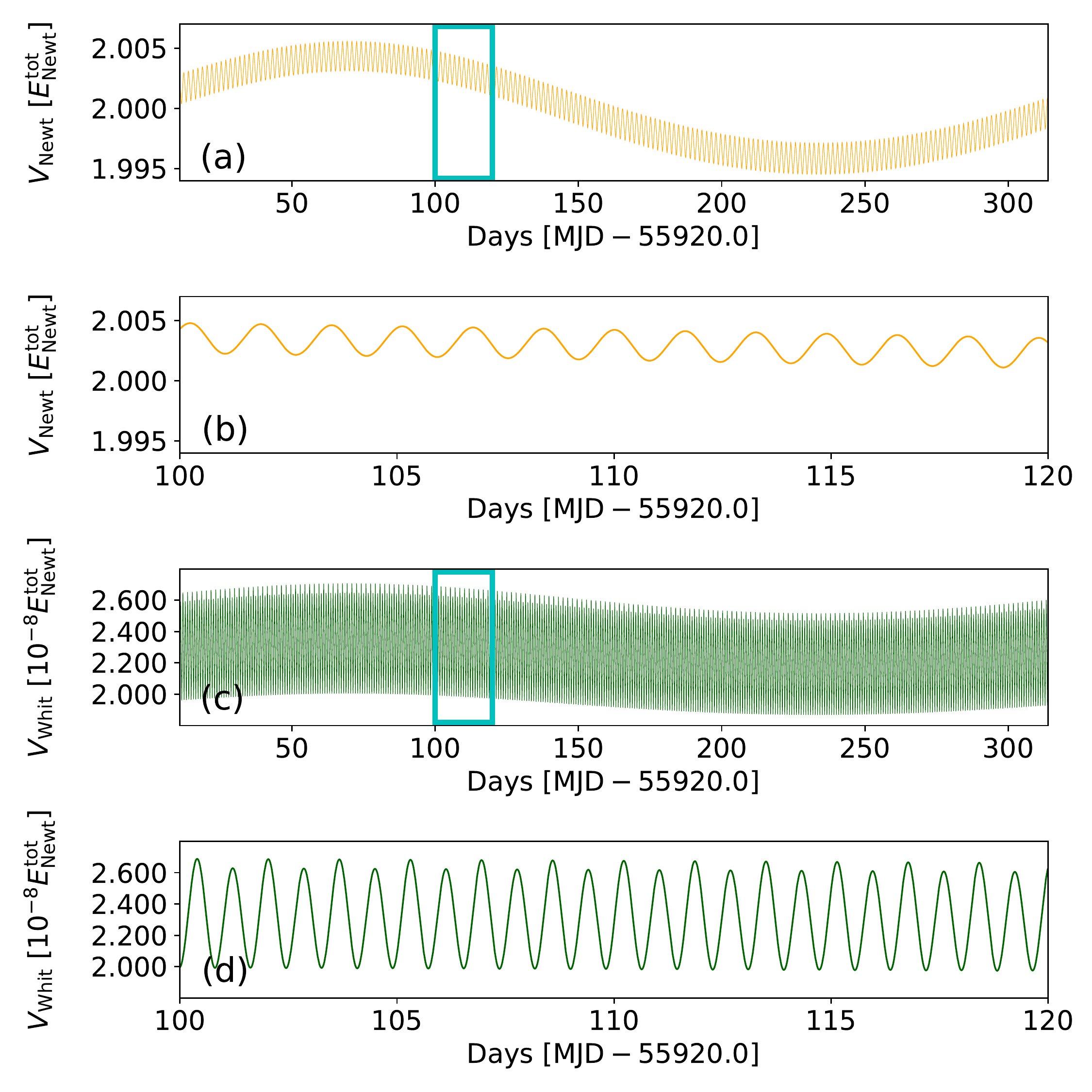}
  \caption{Panels (a) and (c) show respectively the Newtonian and the
    Whitehead potentials versus time for the triple pulsar system. The time
    span is the same as that of Figure~\ref{fig:orbit3d}, namely MJD
    55920.0--56233.9. Panels (b) and (d) show a magnified view of the framed
  regions in panels (a) and (c) respectively.
  \label{fig:potential}}
\end{figure}
%-------------------------------------------------------------------------------

In Figure~\ref{fig:potential}, we plot the Newtonian potential, $V_{\rm Newt}
\equiv \sum_{i \neq j} - \frac{G m_i m_j}{2r_{ij}}$, and the Whitehead
potential, $V_{\rm Whit} \equiv V_\xi (\xi=1)$, in the unit of total Newtonian
orbital energy, $E^{\rm tot}_{\rm Newt} \equiv V_{\rm Newt} + \sum_i
\frac{1}{2} m_i v_i^2$ for the triple system. The total
Newtonian orbital energy $E^{\rm tot}_{\rm Newt} \simeq -8.3 \times 10^{46} \,
{\rm erg}$ is conserved if the gravitational dissipation is ignored. According
to the virial theorem, one has $\langle V_{\rm Newt} \rangle = 2\langle E^{\rm
tot}_{\rm Newt} \rangle$ for Newtonian gravity, where $\langle \cdot \rangle$
denotes an average over time.  As expected, $V_{\rm Newt}$ oscillates with a
short timescale, $P_{\rm b,I}$, and the oscillation is modulated with a large
timescale, $P_{\rm b,O}$. In contrast, $V_{\rm Whit}$ oscillates around its
average value, $\langle V_{\rm Whit} \rangle \simeq -2.0 \times 10^{39}\,{\rm
erg}$, with a short timescale, $\frac{1}{2} P_{\rm b,I}$. A faster oscillation
for $V_{\rm Whit}$ can be understood based on the facts that, the dominant
contribution to $V_{\rm Newt}$ comes from $- \frac{G m_0 m_1}{r_{01}} \propto
\frac{1}{r_{01}}$, while the dominant contributions to $V_{\rm Whit}$ come from
the first and the third terms in the parentheses in
Eq.~(\ref{eq:xi:potential}), both behaving as $\propto \frac{1}{r_{01}^2}$.
There is less long-term {\it modulation} for $V_{\rm Whit}$, compared with
    that of $V_{\rm Newt}$, at the
  timescale of the outer orbit [see panels (a) and (c) in
  Figure~\ref{fig:potential}].  The relative smallness in the modulation of
  $V_{\rm Whit}$ is understood that $V_{\rm Whit}$ is at the first
  post-Newtonian order which receives a smaller contribution from a wider
  orbit, by roughly a factor of $\left({\cal V}_{\rm O}^2/c^2\right) /
  \left({\cal V}_{\rm I}^2/c^2\right) =  {\cal V}_{\rm O}^2 / {\cal V}_{\rm
  I}^2 \simeq 0.03 $ relative to the Newtonian order, where ${\cal V}_{\rm I}$
  and ${\cal V}_{\rm O}$ are the characteristic {\it relative} velocities of
  the inner and the outer orbits.

Speaking of equations of motion for the triple system, for body $i \in \left\{
0, 1, 2 \right\}$ one has an extra acceleration term,
\begin{equation}
  \label{eq:eom}
  \delta\bm{a}^i_{\xi} = - \frac{1}{m_i} \bm{\nabla}^i V_\xi\left( \bm{r}_0,
  \bm{r}_1, \bm{r}_2 \right) \,,
\end{equation}
in addition to the acceleration in the LPI-invariant gravity theory. In
Eq.~(\ref{eq:eom}), $\bm{\nabla}^i$ is the gradient operator for body $i$ with
its coordinate vector $\bm{r}_i$; no summation is assumed for the {\it body
index} $i$ in Eq.~(\ref{eq:eom}). The explicit expressions of
$\delta\bm{a}^i_{\xi}$ are tedious yet not inspiring, therefore, we do not give
them here.

%-------------------------------------------------------------------------------
\section{Mock simulation and parameter estimation}
\label{sec:pe}
%-------------------------------------------------------------------------------

The timing data presented in Ref.~\cite{rsa+14} include radio observations with
the GBT, the Arecibo telescope, and the Westerbork Synthesis Radio Telescope
(WSRT), spanning from MJD 55930.9 to 56436.5 ($\sim1.4$\,yr, or
$\sim4\times10^7$\,s).  Within such a duration, the inner and outer orbits
revolve $\sim300$ and $\sim1.5$ cycles, respectively, while the pulsar spins
$\sim16$ billion cycles.  The timing solution in Ref.~\cite{rsa+14} (see
  the second column in Table~\ref{tab:psr}) is derived with $N^{(0)}_{\rm
  TOA}=26280$ TOAs and results in a weighted root mean squared residual
  $\sigma^{(0)}_{\rm TOA}=1.34\,\mu$s.

Since the TOA data are not publicly available, we simulate mock TOAs closely
following the observational characteristics, as was done in Ref.~\cite{sha16}.
It is well known that there is no general analytical solution for the three-body
problem in gravity given by algebraic expressions and integrals
\cite{bru87,poi92}. To produce mock TOAs, we resort to a machine-precision
$N$-body numerical integrator developed by Rein and his collaborators, which is
implemented in the {\sc rebound}
package\footnote{\url{https://github.com/hannorein/rebound}} \cite{rl12}.
Specifically, we use a 15th-order integrator basing on the Gau\ss{}-Radau
quadrature, {\sc ias15}, that uses adaptive time stepping, and keeps systematic
errors well below machine precision over $10^9$ orbits \cite{rs15}. Such a
precision meets the requirement posed by the accurate pulsar timing data
\cite{sha16}.

For all simulations, we use the parameters of \psr{} reported in
 Ref.~\cite{rsa+14} (see Table~\ref{tab:psr}).
Initial conditions for the system are worked out for MJD 55920.0 which is the
reference epoch for all parameters.  The LPI-violating modification in
Eq.~(\ref{eq:eom}) is augmented to the {\sc ias15} integrator \cite{rs15}.  We
evolve the system in 3-dimensional space for a longer time than the
observational span and cut data keeping the part that corresponds to the real
observation. A spindown model for the pulsar is adopted with $f(t) = f_0 + \dot
f t$ where $t$ is the coordinate time.  The R\"omer delay is obtained by
projecting the 3-dimensional orbit of the pulsar onto the line of sight to the
Earth \cite{tay92}. Relativistic delays, e.g. the Einstein delay and the
Shapiro delay, are ignored, due to the fact that they are not observable in
\psr{} yet \cite{rsa+14}. An exception is the transverse Doppler delay due to
the cross term of the velocities of two orbits, which is approximated for
\psr{} as $R(t) \simeq \frac{1}{c^2} \bm{x}_{\rm I} (t) \cdot \bm{v}_{\rm
O}(t)$, where $\bm{x}_{\rm I}(t)$ and $\bm{v}_{\rm O}(t)$ are respectively the
position vector of the inner orbit and the velocity vector of the outer orbit.
It was shown in Ref.~\cite{sha16} that indeed it is a reasonable approximation.
The end product of one integration is TOAs in the form of $N(t)$ with $N$ the
counting number of pulses.

As mentioned before, $N^{(0)}_{\rm TOA}=26280$ TOAs were collected with
$\sigma^{(0)}_{\rm TOA}=1.34\,\mu$s. Due to the computational cost in MCMC runs
(see below), in our mock simulations we rescale it by a factor of $9$ to $N_{\rm
TOA} = N^{(0)}_{\rm TOA} / 9 = 2920$ and $\sigma_{\rm TOA} = \sigma^{(0)}_{\rm
TOA} / \sqrt{9} \simeq 0.447\mu$s. Such a rescaling is reasonable because on
average it still keeps about 10\,TOAs per inner orbit.

Two strategies are used to sample TOAs \cite{sha16},
\begin{itemize}
  \item {\it uniform sampling}: $N_{\rm TOA} = 2920$ TOAs are generated
    uniformly in time;
  \item {\it step sampling}: $N_{\rm TOA} = 2920$ TOAs are generated with fake
    observing blocks once per week within which TOAs are separated by
    10\,seconds.
\end{itemize}
We consider the {\it step sampling} method more closely resembles the real
observation, however, the sensitivities to LPI violation  from two methods  are
extremely consistent (see below). In total, we simulate five noise realizations
for each method, named as ``{\sf TOA.}$k$'' with $k \in \left\{ 0,1,2,3,4
\right\}$ for {\it uniform sampling} and $k \in \left\{ 5,6,7,8,9 \right\}$ for
{\it step sampling}. Noises in TOAs are generated according to a Gaussian
  random number generator ${\cal N}(0, \sigma_{\rm TOA}^2)$, and they are added
to the ``noiseless'' TOAs to obtain mock TOAs.  For different set of mock TOAs,
the noise generation is independent to each other.

Mock TOAs are generated with the Whitehead's parameter $\xi=0$, with noises
$\sigma_{\rm TOA} \simeq 0.447\mu$s added homogeneously in the uncorrelated
Gaussian form. We want to estimate the extent of these TOAs in constraining
$\xi$. To achieve this task, following Refs.~\cite{rsa+14,sha16}, MCMC runs are
set up to estimate 17 parameters, $\bm{\theta} \equiv
\bm{\theta}_{\rm spin} \cap \bm{\theta}_{\rm orbit} \cap \bm{\theta}_\xi$,
simultaneously in the model.  Parameters in $\bm{\theta}$ include 2 spindown
parameters in $\bm{\theta}_{\rm spin}$, 14 orbital parameters in
$\bm{\theta}_{\rm orbit}$ (see Table 1 for
definition of symbols), and 1 LPI-violating parameter in $ \bm{\theta}_\xi$,
\begin{eqnarray}
  \bm{\theta}_{\rm spin} &\equiv& \left\{ f_0; \, \dot f \right\} \,, \\
  \bm{\theta}_{\rm orbit} &\equiv& \left\{ \left( a \sin i \right)_{\rm I};\,
  P_{\rm b,I};\, \epsilon_{\rm 1,I};\, \epsilon_{\rm 2,I};\, T_{\rm asc,I};\,
   \left( a \sin i \right)_{\rm O};\, P_{\rm b,O};\, \epsilon_{\rm
  1,O};\, \epsilon_{\rm 2,O};\, T_{\rm asc,O};\, \nonumber \right. \\
  && \left. \left( a\cos i \right)_{\rm I};\, \left( a\cos i \right)_{\rm O};\,
  q_{\rm I};\, \delta\Omega \right\} \,, \\
  \bm{\theta}_\xi &\equiv& \left\{ \xi \right\} \,.
\end{eqnarray}
The {\sc Python} package of an affine-invariant MCMC ensemble sampler
\cite{gw10,fhlg13}, {\sc emcee}\footnote{\url{http://dan.iel.fm/emcee}}, is
used to sample the 17-dimensional $\bm{\theta}$-space.  This algorithm has
better performance over traditional MCMC sampling methods (e.g., the
traditional Metropolis-Hasting method), as measured by the smaller
autocorrelation time and fewer hand-tuning input parameters. It transforms the
sampling of the parameter space by an affine transformation such that the
internal algorithm samples an isotropic density, and the efficiency is not
limited by possibly large covariances among parameters \cite{gw10,fhlg13}.
We use uniform priors for all parameters in $\bm{\theta}$, and choose good
starting values to reduce computational cost. Convergence tests are performed
in post-processing to ensure that the starting values do not influence our
parameter estimation. The ranges of parameters are not limited, thus in
principle they can take any values as long as they have support from the
likelihood. For each mock TOA dataset, noiseless template TOAs are generated
on the fly at every MCMC step according to $\bm{\theta}$ that is sampled by the
kernel. These noiseless {\it template TOAs} are compared with the noisy {\it
mock TOAs}, namely ``{\sf TOA.}$k$'' with $k \in \left\{ 0, 1, \cdots, 9
\right\}$. The kernel proceeds the sampling of $\bm{\theta}$ based on the
difference between {\it template TOAs} and {\it mock TOAs}, characterized by
$\chi^2(\bm{\theta})$. With this setting, the
  posterior is directly proportional to $e^{-\chi^2(\bm{\theta})/2}$.

%-------------------------------------------------------------------------------
\begin{figure}
  \includegraphics[width=16cm,center]{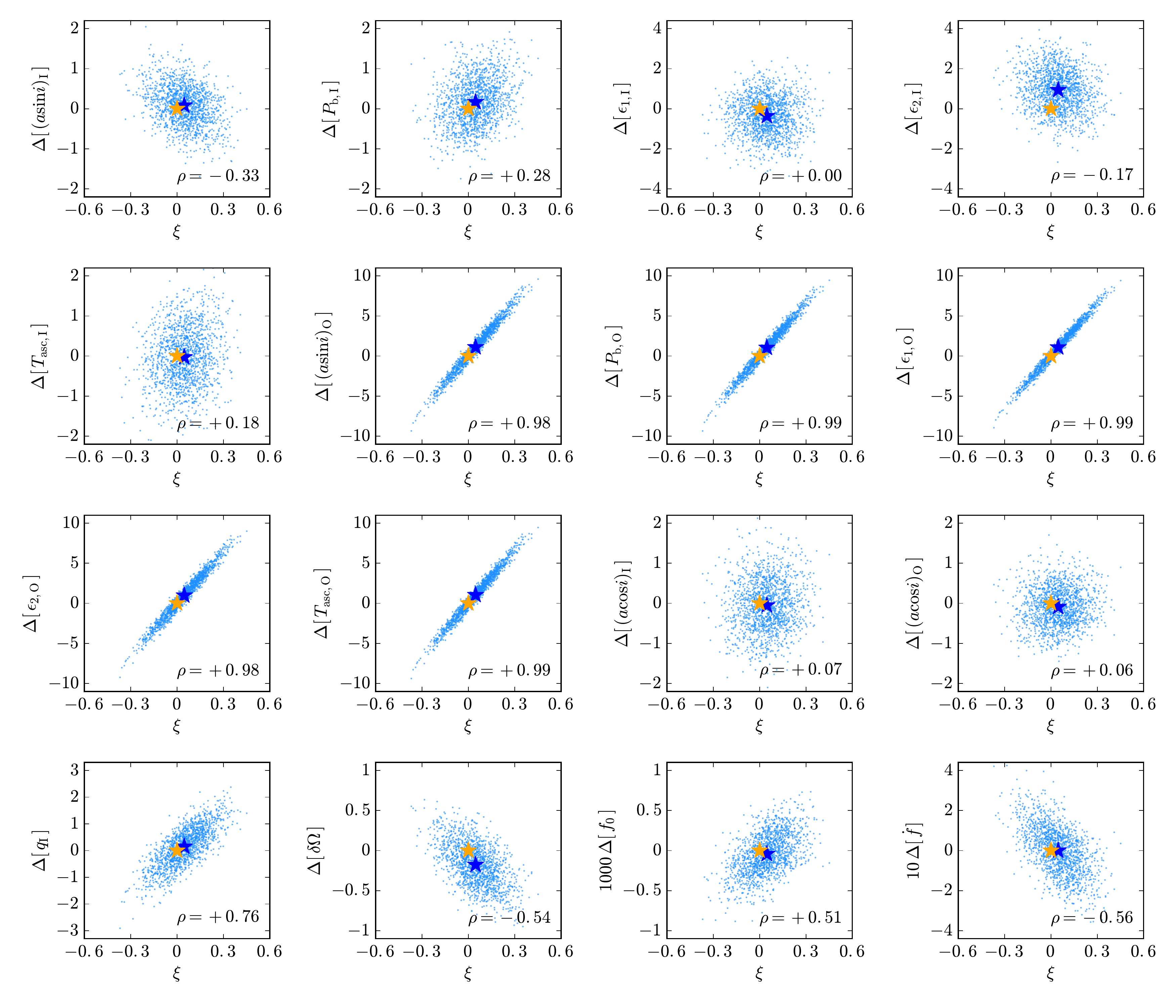}
  \caption{Correlations between $\xi$ and the other 16 parameters from the
    parameter estimation of the mock dataset ``{\sf TOA}.$3$''. The quantity
    $\Delta[{\cal X}] \equiv \left( {\cal X} - {\cal X}^{(0)} \right) /
    \sigma_{\cal X}^{(0)}$, where $ {\cal X}^{(0)}$ and $\sigma_{\cal X}^{(0)}$
    are the values and uncertainties reported in Ref.~\cite{rsa+14} (see
    Table~\ref{tab:psr}).  Orange and
    blue stars are the {\it injected} and {\it recovered} values. The
    correlation coefficients are given at the bottom right corner in each panel.
    Only 1\% of MCMC samples are shown for clarity.
    \label{fig:corr}}
\end{figure}
%-------------------------------------------------------------------------------

%-------------------------------------------------------------------------------
\begin{figure}
  \includegraphics[width=13cm,right]{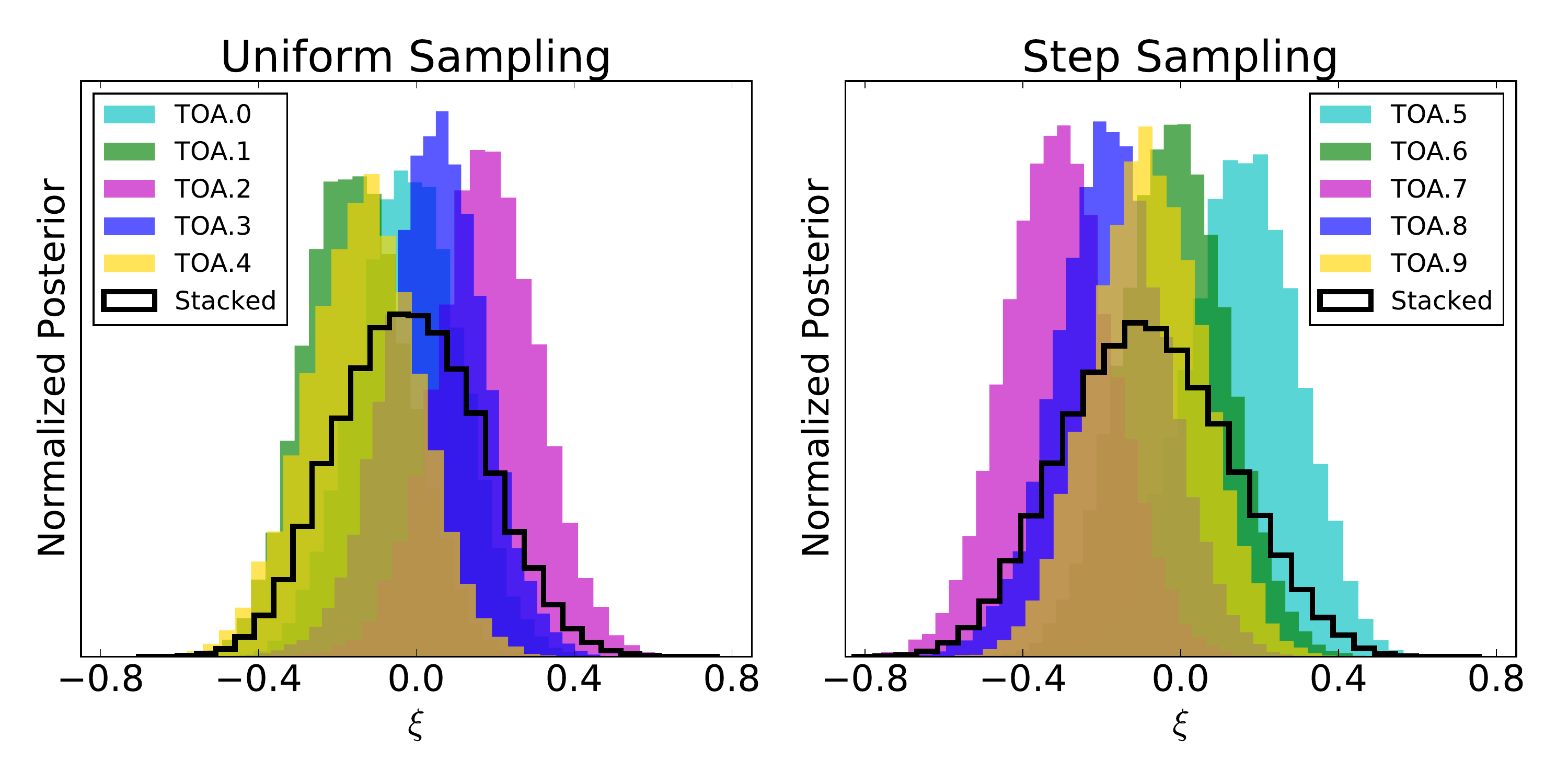}
  \caption{Marginalized and normalized 1-dimensional posterior densities for
    $\xi$ from the parameter estimation of mock datasets with {\it uniform
    sampling} (left) and {\it step sampling} (right). The {\it stacked}
    posterior densities simply stack the posterior samples from different noise
    realizations with the same weight.  \label{fig:posterior}}
\end{figure}
%-------------------------------------------------------------------------------

In total, we have 10 MCMC runs for two sampling methods each with five different
noise realizations. With {\sc emcee}, 44 walkers are adopted for each run, and
44 chains for each set of mock TOAs are accumulated as the end product of MCMC
runs for post-processing.  There are about $300000$ posterior samples for each
mock dataset, and the first half of them are discarded as the {\sc burn-in}
phase \cite{bgjm11}. The Gelman-Rubin statistic is used to assess the
convergence of different chains \cite{gr92}, which tells that the runs have
forgotten the starting values and are in equilibrium.

As was already demonstrated in Ref.~\cite{sha16}, if the {\it template TOAs}
use GR dynamics, the mock TOAs can reproduce the observational {\it
uncertainties} of all orbital parameters in $\bm{\theta}_{\rm orbit}$ within a
factor of two for 13 parameters and a factor of three for the remaining one,
while they underestimate the {\it uncertainties} of $\bm{\theta}_{\rm spin}$.
In the current case, the recovering template is a LPI-violating template with
one extra degree of freedom, therefore, we expect to produce larger
uncertainties, at least for some variables that are strongly correlated with
$\xi$. This is indeed the case, as shown in Figure~\ref{fig:corr} for the
dataset ``{\sf TOA.}$3$'' as an example. We see that $\xi$ strongly correlates
with orbital parameters that pertain to the outer orbit with correlation
coefficients $\rho = 0.98$--$0.99$, while it has relatively smaller
correlations with the other parameters. The correlation with the parameters of
the outer orbit makes the (marginalized) uncertainties of these parameters
(see the third column of Table~\ref{tab:psr}) larger than what was
reported in Ref.~\cite{rsa+14} where the recovering template is LPI-invariant.
The reason for large correlation, we suspect, is that the observational span
only covers about $1.5$ cycles for the outer orbit, which makes parameter
estimation for these elements rather uncertain.\footnote{Ideally it will be
  rather rewarding to study the parameter-estimation problem here with longer
  mock datasets, say, with $\gtrsim 6$ outer orbits, to investigate the
  reduction in the correlations between $\xi$ and the outer orbital parameters,
  and the improvement in constraining $\xi$. However, currently we are limited
  by the speed of the three-body integration and the high dimensionality of the
  parameter space, thus a high computational cost in the MCMC runs. We hope the
code can be speeded up in the future, and address this important question.}
Worthy to mention that, in Figure~\ref{fig:corr} the parameter estimation
recovers the {\it injected} parameters very well, as marked with stars. The
parameter-estimation chains with other mock datasets, ``{\sf TOA.}$k$'' ($k\neq
3$), have similar results.

In Figure~\ref{fig:posterior}, the marginalized 1-dimensional posterior
densities of $\xi$ for all 10 runs are given by normalized histograms. We can
see that though with different sampling strategies and different noise
realizations, the posteriors on $\xi$ are rather consistent, in terms of their
means and variances. In real data, unlike the cases in our
mock datasets where several noise realizations are simulated, only one noise
realization is working and we only have one dataset. To
make a conservative estimate on the expected constraint on $\xi$ from \psr{}, we
stack the posterior samples in each sampling method, and it is shown with black
histograms in Figure~\ref{fig:posterior}.  From the stacked posterior densities,
we obtain a conservative sensitivity of \psr{} in probing the LPI of gravity,
\begin{equation}
  \left| \xi \right| \lesssim 0.4 \,, \quad\quad\quad \mbox{(95\% CL)}\,,
  \label{eq:limit}
\end{equation}
for both sampling methods.

%-------------------------------------------------------------------------------
\section{Discussion}
\label{sec:dis}
%-------------------------------------------------------------------------------

Besides the generic value of $\xi$ in the PPN
framework \cite{wil93,wil14a}, and the specific example of Whitehead's gravity
theory ($\xi=1$) \cite{whi22}, a class of theories called ``quasilinear''
theories of gravity could have a nonzero $\xi$ \cite{wil73}. In these theories,
the PPN parameter $\beta$, that measures the nonlinearity in the superposition
law for gravity, equals to $\xi$; therefore the limit on $\xi$ can be cast as a
limit on $\beta$ as well in these theories. Furthermore, as noted in
Ref.~\cite{sw13}, the constraint on $\xi$ might also limit parameters in the
{\it anisotropic} PPN framework of Ref.~\cite{nor76}, and in the gravity sector
of  SME
\cite{kos04,bk06,kt11,sha14,sha14a}. If the code can be speeded up
  significantly thus the computational cost can be reduced significantly in the
future, similar analysis with multiple non-GR parameters will also be possible
(for example, in the SME framework).

Compared with previous observational constraints on the Whitehead's
parameter~\cite{wil73,wg76,nor87,swk15,sw13}, the expected constraint in
Eq.~(\ref{eq:limit}) is worse than the current
best limit from solitary pulsars~\cite{sw13} by orders of
magnitude.\footnote{If the limit on $\xi$ is converted to a limit on the
  anisotropy of the gravitational constant~\cite{wil93, sw13}, it is worse than
  the current best limit from solitary pulsars~\cite{sw13} by the same orders
of magnitude.} Nevertheless, it has its own virtue.  Firstly, it is a totally
{\it independent} limit yet it is able to rule out the Whitehead's gravity
theory \cite{whi22} alone, that adds to the ``multiple deaths''~\cite{gw08} of
that theory. Secondly, previous limits generally involve extra assumptions, for
examples, the alignment of the Solar spin with the angular orbital momentum of
the Solar system five billion years ago \cite{nor87,sw13} or the statistical
assumptions of unknown angles in the cases of binary pulsars \cite{swk15} and
solitary pulsars \cite{sw13}. The test proposed here is immune to extra
assumptions. Thirdly, compared with the limits from gravimeters on Earth
\cite{wil73,wg76} and the lunar laser ranging experiment \cite{wtb12}, here we
have a strongly self-gravitating body involved.  For some gravity theories, for
example the scalar-tensor gravity \cite{de93,ssb+17}, strong fields will
amplify deviations from GR nonperturbatively.  Even in the case of a
perturbative expansion in the compactness ${\cal C}$, $\xi$ might have a linear
dependence on ${\cal C}$, as in the case of the Nordtvedt parameter
\cite{nor68a,wil93}. If strong fields are relevant, then the limit in
Eq.~(\ref{eq:limit}) could have a large relative merit over weak-field ones due
to the large compactness of the NS, ${\cal C}_{\rm NS} \simeq 0.1$. Fourthly,
compared with experiments that use the Milky Way -- binary-pulsar systems, here
it is a {\it dynamical} three-body system where the third body (the outer WD)
reacts to the gravitational dynamics. Although it is not clear yet but we
suspect that this might have a standing for some specific gravity theories.

The estimation in Eq.~(\ref{eq:limit}) is obtained with mock data simulated
closely following the real observation for a time span about $1.4$\,yr. In
reality, new observations have accumulated more data with probably better
qualities.  Up to the time of writing, about 6 outer orbits are covered,
compared with $\sim1.5$\,orbits used in the simulation. This will vastly break
parameter degeneracy and reduce the strong correlations ($\rho=0.98$--$0.99$)
seen in Figure~\ref{fig:corr} with the elements of the outer orbit. Thus these
new data are expected to give an even tighter limit than that by a naive
rescaling.  In future, new radio telescopes like the Five-hundred-meter Aperture
Spherical Telescope (FAST)~\cite{nlj+11} and the Square Kilometre Array (SKA)
\cite{kbc+04,ssa+15} will provide better sensitivities in obtaining TOAs for the
triple system.  Moreover, although triple pulsars are rare, there is a chance of
hosting about 100 such systems in the Milky Way \cite{rsa+14}, and the SKA is
going to discover almost all of them \cite{kbk+15}. If an even tighter triple
system is discovered, a better test of LPI in gravity could be conducted.

Lastly, we want to stress that, although our mock TOAs are able to reproduce
major features of the observation in Ref.~\cite{rsa+14}, they are nevertheless
simplified compared with the complications in real data, e.g., the
heteroscedasticity in TOAs from different telescopes (the GBT, the Arecibo
telescope, and the WSRT), the removal of (probably time-dependent) interstellar
dispersion, the irregular jumps between different observing sessions, and so
on.  Also, there will be correlations of parameters with the parallax and the
proper motion of \psr{}, which could be resolved with the Very Long Baseline
Array (VLBA) \cite{rsa+14}. We are not expecting that simulated mock data
to cover all these observational facts. Nevertheless, we believe that the
simulations have include major features of the triple system.  An analysis with
the real data could settle the result firmly. We hope the analysis done here
will simulate observers to analyze the real data to test LPI.

%-------------------------------------------------------------------------------
\subsection*{Acknowledgements}

The author thanks Roland Haas and Hanno Rein for help with codes, and Vivien
Raymond for discussion on statistics.  The Markov-chain Monte Carlo runs were
performed on the {\sc vulcan} cluster at the Albert Einstein Institute in
Potsdam-Golm.
%-------------------------------------------------------------------------------

\section*{References}

%\bibliography{/Users/lshao/shao/reference.bib}{}
%\bibliographystyle{unsrt}

%---------------------------------------------------------------------
\begin{landscape}
\begin{table}[htp]
  \centering
  \caption{\label{tab:psr} Parameters for the spindown of the pulsar and the
  orbits of the triple system. Values in the second column are from
  Ref.~\cite{rsa+14}, while those in the third column are from the parameter
  estimation on the mock dataset ``{\sf TOA}.$3$'' while allowing a nonzero
whitehead parameter in the MCMC runs. Parenthesized numbers represent the
1-$\sigma$ uncertainty in the last digits quoted. }
  \begin{tabular}{lll}
    \hline\hline
    Parameter & Observation & Simulation \\
    \hline
    \multicolumn{3}{l}{\sc Spindown parameters} \\
    \hline
    Pulsar spin frequency, $f_0$ & $365.953363096(11)$\,Hz &
    $365.953363095999(3)$\,Hz \\
    Spin frequency derivative, $\dot f$ & $-2.3658(12) \times 10^{-15}\,{\rm
    Hz\,s}^{-1}$ & $-2.3658(1) \times 10^{-15} \,{\rm
    Hz\,s}^{-1}$ \\
    \hline
    \multicolumn{3}{l}{\sc Inner Keplerian parameters for pulsar orbit} \\
    \hline
    Semimajor axis projected along line of sight, $\left( a\sin i \right)_{\rm
    I}$ & $1.21752844(4)$\,ls & $1.21752844(2)$\,ls \\
    Orbital period, $P_{\rm b,I}$ & $1.629401788(5)$\,d & $1.629401789(3)$
    \,d \\
    Eccentricity parameter, $\epsilon_{\rm 1,I} \equiv \left( e \sin \omega
    \right)_{\rm I}$ & $6.8567(2)\times 10^{-4}$ & $6.8566(2)\times 10^{-4}$ \\
    Eccentricity parameter, $\epsilon_{\rm 2,I} \equiv \left( e \cos \omega
    \right)_{\rm I}$ & $-9.171(2)\times10^{-5}$ & $-9.169(2)\times10^{-5}$ \\
    Time of ascending node, $T_{\rm asc,I}$ & MJD\,$55920.407717436(17)$ &
    MJD\,$55920.40771744(1)$ \\
    \hline
    \multicolumn{3}{l}{\sc Outer Keplerian parameters for center of mass of
    inner binary} \\
    \hline
    Semimajor axis projected along line of sight, $\left( a\sin i \right)_{\rm
    O}$ & $74.6727101(8)$\,ls & $74.672711(2)$\,ls \\
    Orbital period, $P_{\rm b,O}$ & $327.257541(7)$\,d & $327.25755(2)$\,d \\
    Eccentricity parameter, $\epsilon_{\rm 1,O} \equiv \left( e \sin \omega
    \right)_{\rm O}$ & $3.5186279(3)\times 10^{-2}$ & $3.5186282(8)\times 10^{-2}$ \\
    Eccentricity parameter, $\epsilon_{\rm 2,O} \equiv \left( e \cos \omega
    \right)_{\rm O}$ & $-3.462131(11)\times10^{-3}$ & $-3.46212(3)\times10^{-3}$  \\
    Time of ascending node, $T_{\rm asc,O}$ & MJD\,$56233.935815(7)$ &
    MJD\,$56233.93582(2)$ \\
    \hline
    \multicolumn{3}{l}{\sc Interaction parameters} \\
    \hline
    Semimajor axis projected in plane of sky, $\left( a\cos i \right)_{\rm I}$
    & $1.4900(5)$\,ls & $1.4900(3)$\,ls \\
    Semimajor axis projected in plane of sky, $\left( a\cos i \right)_{\rm O}$
    & $91.42(4)$\,ls & $91.42(2)$\,ls \\
    Ratio of inner companion mass to pulsar mass, $q_{\rm I} \equiv m_{\rm
    WD,I} / m_{\rm NS}$ & $0.13737(4)$ & $0.13738(3)$ \\
    Difference in longitudes of ascending nodes, $\delta\Omega$ & $2.7(6)
    \times 10^{-3}$\,deg & $2.6(2) \times 10^{-3}$\,deg \\
    \hline
  \end{tabular}
\end{table}
\end{landscape}
%---------------------------------------------------------------------

\end{document}